\documentclass[twocolumn, a4paper, times, trackchanges]{aastex61}
\usepackage[english]{babel}
\usepackage[utf8x]{inputenc}
\usepackage{amsmath}
\usepackage{graphicx}
\usepackage{natbib}

\usepackage{chngcntr}
\begin{document}
\title{Evolution of the solar Ly-$\alpha$ line profile during the solar cycle. II. How accurate is the present radiation pressure paradigm for interstellar neutral H in the heliosphere?}

\correspondingauthor{Izabela Kowalska-Leszczynska}
\email{ikowalska@cbk.waw.pl}

\author{Izabela Kowalska-Leszczynska}
\affiliation{Space Research Centre PAS (CBK PAN),\\
Bartycka 18A, 00-716 Warsaw, Poland}

\author{Maciej Bzowski}
\affiliation{Space Research Centre PAS (CBK PAN),\\
Bartycka 18A, 00-716 Warsaw, Poland}

\author{Justyna M. Sok{\'o}{\l}}
\affiliation{Space Research Centre PAS (CBK PAN),\\
Bartycka 18A, 00-716 Warsaw, Poland}

\author{Marzena A. Kubiak}
\affiliation{Space Research Centre PAS (CBK PAN),\\
Bartycka 18A, 00-716 Warsaw, Poland}

\begin{abstract}
Following the derivation of a more accurate model of the evolution of the solar Lyman-$\alpha$ line with the changing solar activity by \citet{kowalska-leszczynska_etal:18a} (IKL18) than the formerly used model by \citet{tarnopolski_bzowski:09} (ST09), we investigate potential consequences that adoption of the resulting refined model of radiation pressure has for the model distribution of interstellar neutral (ISN) H in the inner heliosphere and on the interpretation of selected observations. We simulated the ISN H densities using the two alternative radiation pressure models and identical models of all other factors affecting the ISN H distribution. We found that during most of the solar cycle, the IKL18 model predicts larger densities of ISN H and PUIs than ST09 in the inner heliosphere, especially in the downwind hemisphere. However, the density of ISN H at the termination shock estimated by \citet{bzowski_etal:08a} obtained using ST09 does not need revision, and the detection of ISN D by IBEX is supported. However, we point out the existence of a considerable absorption of a portion of the solar Lyman-$\alpha$ spectral flux inside the heliosphere. Therefore, the model of radiation pressure for ISN H is still likely to need revision, and hence the available models of ISN H are not self-consistent.
\end{abstract}

\section{Introduction}
\label{sec:intro}
Interstellar neutral hydrogen (ISN H) is the dominant component of the Local Interstellar Medium (LISM) surrounding the heliosphere \citep{bzowski_etal:09a}. Since the Sun is moving with respect to the LISM at $\sim 25.7$~km~s$^{-1}$ \citep{bzowski_etal:14a, bzowski_etal:15a, mccomas_etal:15a, schwadron_etal:15a}, ISN H is able to penetrate inside the heliopause and reach 1~AU. Measurements of ISN H, both the primary and secondary populations, as well as measurements of the derivative populations (pickup ions -- PUIs, energetic neutral atoms -- ENAs, heliospheric resonant backscatter glow) brought information on important aspects of the LISM, including the density of ISN H \citep{bzowski_etal:08a} and the deflection of the ISN H flow direction from the direction of Sun's motion through the LISM due to distortion of the heliosphere by the interstellar magnetic field \citep{lallement_etal:05a}, as well as on the evolution of the solar wind structure during the solar cycle \citep[e.g.,][]{bzowski_etal:03a, lallement_etal:10b}.

The primordial (``primary'') ISN H is heavily processed within the heliospheric interface \citep{baranov_malama:93, heerikhuisen_etal:06a, heerikhuisen_etal:15a, izmodenov_etal:09a, izmodenov_alexashov:15a} by charge-exchange collisions with the perturbed plasma flowing past the heliopause and also inside the termination shock, due to ionization by charge-exchange with solar wind protons and photoionization. As a result, a secondary population of ISN H is created in the outer heliosheath, which also penetrates inside the heliosphere \citep{baranov_malama:93}. The density, velocity, and thermal spread of the primary and secondary populations of ISN H in the regions where they are measured (i.e., between $\sim 1$ and $\sim 10$~AU) are very sensitive to the intensity of ionization processes and to their variations with time and heliolatitude \citep{rucinski_bzowski:95b, bzowski_etal:97, bzowski_etal:02}. 

Another important factor that shapes the distribution of ISN H density inside the heliosphere is the resonant radiation pressure. The radiation pressure acting on H atoms inside the heliosphere is due to the EUV radiation emitted by the Sun in the chromospheric Lyman-$\alpha$ line. This line features a self-reversed shape with two horns. As a result of this line shape, radiation pressure acting on an individual H atom is a strong function of radial velocity of this atom. 

\begin{figure}
\centering
\includegraphics[width=0.8\columnwidth]{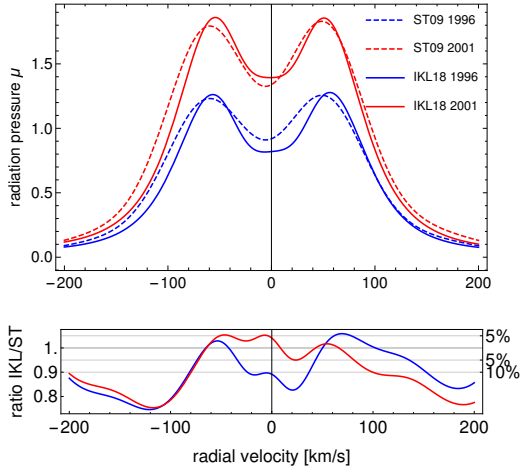}
\caption{ 
%{\em{fig:profile}}
Profile of the Lyman-$\alpha$ solar line in units of radiation pressure (where $\mu=1$ means that radiation pressure compensates the gravity) as a function of radial velocity for the two models (solid lines represent the new model by IKL18 and dashed lines the old model by ST09) and for two phases of solar cycle (red lines correspond to the solar activity maximum in 2001 and blue lines present the profile for the solar activity minimum in 1996). Bottom panel shows ratio between the two models for the two solar activity phases.
\label{fig:profile}
}

\end{figure}

The distributions of density, bulk velocity, and thermal spread of ISN H inside the heliosphere are sensitive to variations of the solar flux in the Lyman-$\alpha$ line and of the line profile during the solar activity cycle \citep{tarnopolski_bzowski:09}. Therefore, it is important to have a precise model of the solar Lyman-$\alpha$ line profile and its modifications due to the varying solar activity. A model of this profile and its variations as a function of the total solar irradiance in the Lyman-$\alpha$  line was proposed by \citet{tarnopolski_bzowski:09} (further on: TB09) based on a limited set of observations from \citet{lemaire_etal:05}. Recently, based on a much more extensive observation set from \citet{lemaire_etal:15a}, \citet{kowalska-leszczynska_etal:18a} (further on: IKL18) proposed a refined functional form of the solar Lyman-$\alpha$ line profile and its variation with the total Lyman-$\alpha$ flux. This latter model is different in some important aspects from the model from TB09, especially for the conditions of low solar activity (see Figure~\ref{fig:profile}). However, the TB09 model was used in several important studies, both modeling and experimental, requiring assessments of the density and other parameters of ISN H in various regions of the heliosphere \citep[e.g.,][]{bzowski_etal:08a, izmodenov_etal:13a, schwadron_etal:13a, fayock_etal:15a, katushkina_etal:15b} and of energetic neutral atoms in the heliosphere \citep[e.g.,][]{bzowski_tarnopolski:06a, bzowski:08a, bzowski_etal:13a, mccomas_etal:10c, mccomas_etal:12c, mccomas_etal:14b, mccomas_etal:17a, swaczyna_etal:16a}. It was also used in the pioneering studies of ISN D in the heliosphere, first modeling \citep{tarnopolski_bzowski:08a, kubiak_etal:13a}, and then experimental \citep{rodriguez_etal:13a, rodriguez_etal:14a}, which resulted in the first direct detection of ISN D inside the heliosphere. 

In this paper we present a comparison of the predictions for the density, velocity, and thermal spread of the primary and secondary populations of ISN H obtained using a state of the art model of the ISN H and PUI densities inside the heliosphere, with radiation pressure based on the TB09 and IKL18 models of the solar Lyman-$\alpha$ line. We are looking for a region in space where the two models of radiation pressure give dissimilar predictions for the ISN H and PUI densities. We also discuss the accuracy of the existing determination of the ISN H density in the LISM based on observations of PUIs on Ulysses. 

Subsequently, we present a detailed comparison of predictions obtained using the radiation pressure models from TB09 and IKL18 for the expected flux, speed, and energy of ISN H atoms impacting the IBEX-Lo detector \citep{fuselier_etal:09b}, which up to now has been the only instrument capable of direct sampling of ISN H \citep{saul_etal:12a, saul_etal:13a}. This aspect is particularly important because of the discovery by \citet{schwadron_etal:13a} and \citet{katushkina_etal:15b} that state of the art models of the heliosphere using the existing models of ionization rate variations and, importantly, of the radiation pressure, cannot reproduce the proportions between the fluxes of ISN H observed by IBEX in different energy bands. The authors of these findings suggested that the most likely reason for this discrepancy is an inadequacy of the presently available models of radiation pressure. 

In this context, we also consider expected modifications of radiation pressure due to absorption of the solar spectral flux by the ISN H gas inside the termination shock. This modification, to our knowledge, has been neglected up to now in heliospheric models. We identify the regions where this absorption is large enough to affect the magnitude of radiation pressure and study differences between these regions predicted by ST09 and IKL18,  but leave an assessment of the effect of this modification to later studies.

\section{Calculations}
\label{sec:calculations}

The calculations used in this study were carried out using the latest version of the numerical strain of the Warsaw Test Particle Model \citep[nWTPM; ][]{tarnopolski_bzowski:09,sokol_etal:15b}. This model is based on the concept proposed by \citet{rucinski_bzowski:95b}, in which the density and higher moments of interstellar gas in a selected moment of time and location inside the heliosphere are obtained from direct numerical integration of the distribution function of this gas, calculated for this time and location. The local distribution function is calculated based on the hot-model paradigm proposed originally by \citet{fahr:78, fahr:79}. However, important refinements include accounting for the following effects, calculated separately for each of the test-particle atoms contributing to the distribution function: 

(1) Variations in the strength of radiation pressure in time \citep{rucinski_bzowski:95b} included by numerical solving the atom's equation of motion, with the force dropping with the square of solar distance but being modulated in time in synchronization with variations of the solar Lyman-$\alpha$ flux.

(2) The magnitude of radiation pressure being a function of atom radial speed (varying along the orbit) due to the Doppler effect \citep{tarnopolski_bzowski:09}.

(3) Variation in the ionization loss rate with time\linebreak
\citep{rucinski_bzowski:95b} due to the time-variation of the ionizing factors (e.g., charge exchange and photoionization).

(4) Variation of the ionization loss rate with heliolatitude, mostly due to the latitudinal variation of the solar wind speed and density \citep{bzowski_etal:01a, bzowski_etal:02}.

(5) Adoption of a model of the local distribution function as due to a superposition of two homogeneous Maxwell-Boltzmann distributions of ISN H beyond the heliopause, representing the primary and secondary populations of ISN H \citep{scherer_etal:99}.

The model was used with the ionization rate obtained using the latest versions of models of the solar wind parameter variation with time and heliolatitude \citep{sokol_etal:13a} and ionization rate \citep{bzowski_etal:12a, bzowski_etal:13a}. The ionization rate and radiation pressure models are based on measurements and therefore we consider them as able to realistically reproduce the actual conditions in the heliosphere, within the uncertainties. In the two comparison simulations used in this paper, the only differences are the solar Lyman-$\alpha$ line models. 

In the simulations, we adopted very similar assumptions to those used by \citet{bzowski_etal:08a} to determine the ISN H density at the termination shock. We assumed that the local distribution function of ISN H inside the heliosphere is a superposition of the primary and secondary populations of ISN H, with the parameters of these populations at the termination shock  close to these obtained by \citet{bzowski_etal:08a}, but slightly modified based on insight obtained from later studies. The speed, inflow direction, and temperature of the primary population was adopted as identical to those found by  \citet{bzowski_etal:15a} based on analysis of direct-sampling observations of ISN He by IBEX. The density of the primary population was taken identical to that found for this population by \citet{bzowski_etal:08a}. For the secondary population, we adopted the density, temperature, and inflow speed identical to these parameters found for the secondary population of ISN H by \citet{bzowski_etal:08a}, but the inflow direction was adopted identical to that found  by \citet{kubiak_etal:16a} for the Warm Breeze, which most likely is the secondary population of ISN He. We note here that the angle between the inflow directions of the primary and secondary populations, equal to $\sim 8\degr$, is very close to the difference between the inflow directions of the primary and secondary populations of ISN H obtained from the Moscow Monte Carlo model of the heliosphere by \citet{izmodenov_alexashov:15a}.
The parameters of the primary and secondary populations are listed in Table \ref{tab:params}.

\begin{deluxetable}{cccc}
\tablecaption{\label{tab:params} Parameters of the primary and secondary populations adopted for this calculations.}
\tablehead{
	\colhead{Parameter description}&	\colhead{Parameter} & \colhead{ Primary} & \colhead{Secondary}
	}
\startdata
Density at termination shock & $n_{TS}$ [cm$^{-3}$] & $0.031$  & $0.054$ \\
Temperature                  & T [K]                & $7443$   & $16300$ \\
Speed of the inflow          & v [km s$^{-1}$]      & $-25.78$ & $-18.74$\\
Ecliptic longitude           & $\lambda$ [$\degr$]    & $255.75$ & $251.57$\\
Ecliptic latitude            & $\beta$ [$\degr$]     & $5.17$  & $11.95$ \\
\enddata
\end{deluxetable}

The calculations were performed in a plane containing the Sun and the upwind-downwind line, with the crosswind directions parallel to the ecliptic plane, on a grid logarithmically spaced in solar distance and uniformly spaced in the azimuthal angle (with a finer pitch in the downwind region). The calculations were carried out for the epochs corresponding to the solar minimum (1996.0) and solar maximum (2001.0) conditions. Additionally, for selected radial lines from the adopted grid, listed in Table~\ref{tab:directions}, we performed comparison simulations for a long time series with a time step of one Carrington rotation period.

Throughout the paper, we show IKL18/ST09 ratios of the following quantities: density, vector flux, and IBEX flux. We seek to identify regions where this ratio does not exceed selected magnitudes, from 1.5 (a $\sim 50$\% difference) down to 5\%. For the discussion of the H and D flux at IBEX (Section~\ref{sec:IBEX}) we had to choose dates from another solar cycle because IBEX was launched in 2008. 

\section{The density and velocity of ISN H and the density of pickup ions}
\label{sec:gasParams}
\subsection{Density}
\begin{figure}
\centering
\includegraphics[width=0.85\columnwidth]{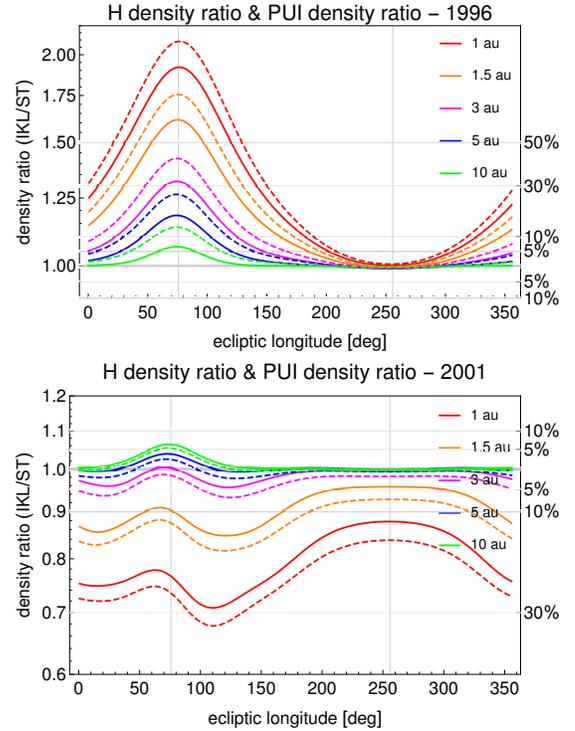}
\caption{
%{\em{fig:density1}}
Ratio of densities calculated using two radiation pressure models (IKL18 and ST09) for ISN H (solid lines) and H$^+$ PUIs (dashed lines), shown as a function of ecliptic longitude. Results are presented for the distances from the Sun color-coded in the figure. Top panel presents the results for the solar minimum conditions (1996), and bottom panel represents the solar maximum conditions (2001). The vertical grids mark the upwind (255.7$\degr$) and downwind (75.7$\degr$) longitudes, while the horizontal grids the differences between the models at 50\%, 30\%, 10\%, and 5\% level (both positive and negative). }
\label{fig:density1}
\end{figure}

\begin{figure}
\centering
\includegraphics[width=0.85\columnwidth]{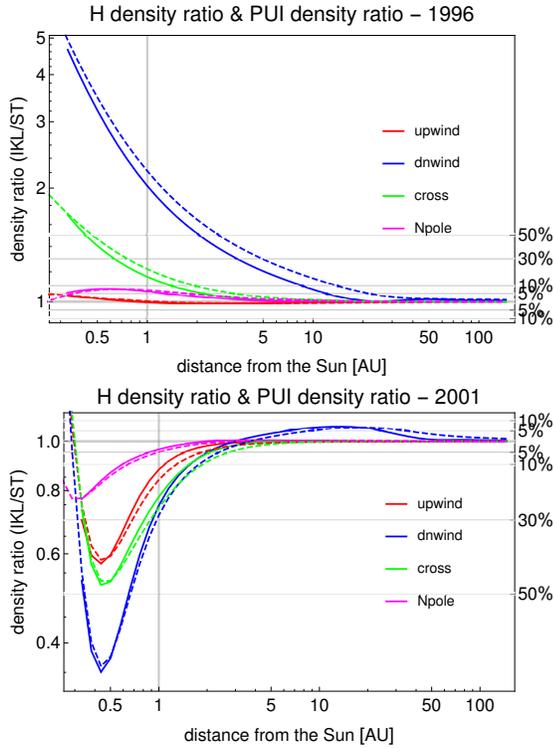}
\caption{
%{\em{fig:density2}}
Ratio of densities of ISN H and PUIs (solid and dashed lines, respectively), calculated using the two radiation pressure models (IKL18 and ST09), shown as a function of distance from the Sun. The ratios are presented for the radial directions defined in Table~\ref{tab:directions}, with color-coding indicated in the figure. Top panel corresponds to the solar minimum (1996), and bottom panel to solar maximum (2001) conditions. 
}
\label{fig:density2}
\end{figure}

We start with a discussion of the density of ISN H. We analyze the ratios IKL18/ST09 of the density calculated using the new radiation pressure model (IKL18) to that based on the old model (ST09) for the solar minimum and maximum conditions along ecliptic rings at selected distances from the Sun (Figure~\ref{fig:density1}) and along the selected directions in space from Table~\ref{tab:directions} (Figure~\ref{fig:density2}). The choice of the directions in our analysis is connected with the geometry of the inflow of interstellar gas. The first three directions in the table (``upwind'', ``dnwind'', and ``cross'') correspond to the upwind, downwind, and one of the crosswind directions near the ecliptic plane, while ``Npole'' corresponds to the heliographic north pole axes.

\begin{figure}
\centering
\includegraphics[width=0.85\columnwidth]{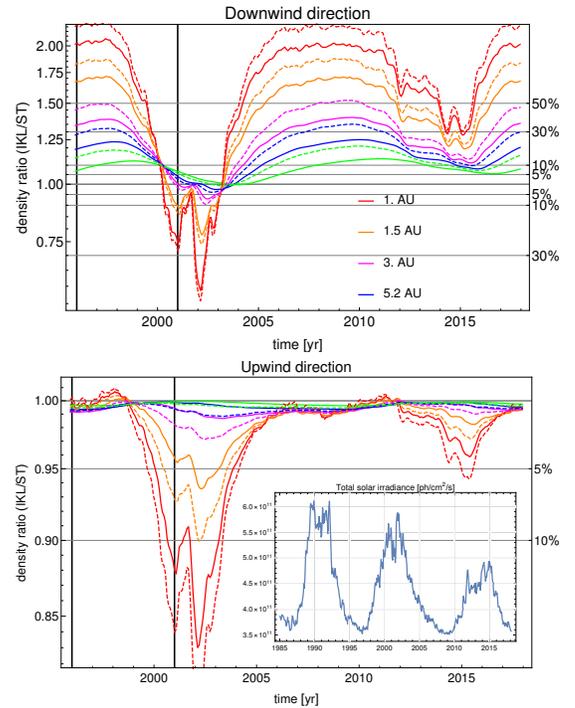}
\caption{
%{\em{fig:density3}} 
Ratios of ISN H and PUI densities calculated using two radiation pressure models (IKL18 and ST09) as a function of time. The results are shown for the distances from the Sun indicated in the panels. Top panel presents downwind direction, while bottom panel presents the upwind direction. Solid lines represent the ISN H density ratios, and the dashed lines the PUI density ratios.
Furthermore as a inset to the bottom panel we show the solar irradiance in the Lyman-$\alpha$ line for the three most recent solar cycles, based on the LASP observations supplemented with proxies \citep{woods_etal:15a}. Note that the maximum/minimum ratio of irradiances for the the maximum of 2015 is approximately half that of the maximum of 2001. The right-hand vertical axis is the percentage of an upward or downward deviation from 1 of the density ratio.
}
\label{fig:density3}
\end{figure}

The density ratio for the minimum of solar activity is presented in the top panel of Figures \ref{fig:density1} and \ref{fig:density2} and those for the solar maximum conditions in the bottom panels of these figures. Additionally for the upwind and downwind directions, in Figure~\ref{fig:density3} we show the variation of the IKL18/ST09 density ratios with time at the same distances as in Figure~\ref{fig:density1}. 

\begin{deluxetable}{ccc}
\tablecaption{\label{tab:directions} Ecliptic coordinates of the directions shown in the plots.}
\tablehead{
		\colhead{direction name} & \colhead{ Ecliptic longitude ($\lambda$)} & \colhead{Ecliptic latitude ($\beta$)}
	}
\startdata
``upwind''  & $255.7$  & $5.17$ \\
``dnwind'' & $75.7$   & $-5.17$ \\
``cross1'' & $165.7$ & $0$\\
%``cross2'' & $345.04$ & $-7.25$ \\
``Npole''  & $346.72$  & $82.75$ \\
%``Spole''  & $166.72$ & $-82.08$\\
\enddata
\end{deluxetable}

The density differences are almost negligible along the upwind axis (within $\sim 5$\% for all distances), but they increase with the offset angle from the upwind direction and reach maximum at the downwind axis. Inspection of Figure~\ref{fig:density1} suggests that a density difference between the two models exists in the entire volume outside a cone centered at the upwind direction. For the 10\% difference level, the half-width of this cone decreases from $150\degr$ at 5~AU to $\sim 70\degr$ at 1~AU during low solar activity. For a high solar activity, as during the solar maximum of 2001, the density difference changes its sign for almost all offset angles from the upwind direction, and within $\sim 1$~AU the entire volume is affected. Due to the relatively low propagation speed of ISN H along the downwind axis, the differences between the two models along the downwind axis have wave-like behavior, as illustrated in the lower panel of Figure~\ref{fig:density2}. 

The ISN H densities obtained using the IKL18 radiation pressure model are larger than those calculated using the ST09 radiation pressure during most of the solar cycle. An exception may be a relatively short interval of time near the maximum of solar activity, if the solar Lyman-$\alpha$ flux is sufficiently strong, as evidenced by the different depths of the minima in Figure~\ref{fig:density3} occurring for the strong solar activity maximum of $\sim 2001$ and the weak maximum of $\sim 2015$. That effect is clearly visible in the total solar irradiance plot that is shown as an inset on Figure \ref{fig:density3}.

A conclusion from this portion of the study is that the ISN H densities calculated using the old model of radiation pressure (ST) within the downwind portion of the heliosphere up to $\sim 7$~AU were inaccurate by at least 10\%, with the discrepancies increasing towards the Sun and towards the downwind axis. 

\subsection{Radial velocity and speed}
\label{sec:speed}
Radial velocity of ISN H inside the heliosphere affects the radiation scattering properties of the gas and consequently the heliospheric backscatter glow. Because of the Doppler effect, the gas that has a non-zero radial velocity absorbs a different portion of the solar spectrum than that for radial velocity equal to 0. Therefore, radiation pressure acting on hydrogen atoms strongly depends on the radial velocity of these atoms. Our simulations suggest that radial velocity is little affected and the bulk velocities of ISN H predicted using the radiation pressure model from ST09 are accurate within $\sim 2$\% in the upwind hemisphere and $\sim 5$\% in the downwind region.

Bulk speed (i.e., the magnitude of the bulk velocity vector) effect is important for direct-sampling measurements. The construction of the IBEX-Lo detector makes it very sensitive to the impact speed \citep{fuselier_etal:09b} of individual atoms on the conversion surface of the detector. Therefore, the accuracy of modeling of the speed of ISN H potentially affects the accuracy of conclusions drawn from analysis of observations.

Differences between model results in bulk speed of ISN H are very small by percentage. In all cases, the model predictions agree within 10\%, i.e., within a few km~s$^{-1}$. During the solar minimum, the largest differences are in the downwind direction, but the maximum difference shifts towards the upwind direction with decreasing distance. The speed predicted by IKL18 is smaller around downwind direction and higher around "upwind" direction than the speeds obtained using ST09. During the solar maximum, the angular structure is much more complicated. We simulate the ISN H inside the heliosphere as a superposition of the primary and secondary populations, and since the gas inside the heliosphere is collisionless, it makes sense to consider the two populations separately. The derived speed of the primary population is larger for the ST09 model. Some differences in the upwind direction exist, with the magnitude decreasing with the distance from the Sun. The secondary population during the solar maximum behaves similarly, but the magnitudes of the difference near the downwind direction are smaller. Again, the magnitude of the differences is relatively small: even the largest of them are at a level of 8\%.

\subsection{Pickup ion flux and density}
\label{sec:puiDensity}
PUIs are former atoms of ISN H that have been ionized inside the heliosphere and intercepted (``picked up'') by the Lorentz force of the magnetic field frozen in the solar wind plasma \citep{vasyliunas_siscoe:76}. They make a separate population within the solar wind plasma that is propagating with the speed practically equal to that of the solar wind. In the scenario with a stationary ionization rate and stationary flow of ISN H, the flux of PUIs $F_{\text{PUI}}(\vec{r})$ at a given location in space $\vec{r}$ can be approximated by 
\begin{equation}
F_{\text{PUI}}(\vec{r}) = \frac{1}{r^2} \int \limits_{r_0}^{r} n_{\text{H}}(\vec{r}') \beta(\vec{r}') r'^2 d r'
\label{eq:PUIFlux1},
\end{equation}
where $r_0$ is the solar radius, $n_{\text{H}}(\vec{r})$ the ISN H density at $\vec{r}$, and $\beta(\vec{r})$ the ionization rate at $\vec{r}$ \citep{vasyliunas_siscoe:76}. The same formula can be adopted for a time-variable scenario providing that the rate of change of the ISN H density at $\vec{r}$ is much lower than the propagation time of solar wind from the Sun to $\vec{r}$ \citep{rucinski_etal:03}. We adopted  Equation~\ref{eq:PUIFlux1} to calculate the PUI flux with the ISN H densities and ionization rates variable in $t$ using the full time-dependent version of the nWTPM model (more in Sok{\'o}{\l} et al. 2018, in preparation).

Since flux can be approximated as a product of density and propagation speed, we calculated the densities of PUIs at $\vec{r}$ and $t$ from the formula
\begin{equation}
n_{\text{PUI}}(\vec{r}, t) = F_{\text{PUI}}(\vec,{r}, t)/v_{\text{SW}}(\vec{r}, t),
\label{eq:PUIDens}
\end{equation}
where $v_{\text{SW}}(\vec{r}, t)$ is the solar wind speed at $\vec{r}$, $t$, assumed to be independent of the solar distance but varying with time and heliolatitude. 

We calculated the PUI flux assuming either IKL18 or ST09 radiation pressure models, using identical ionization rates and solar wind parameters. In the following, we discuss the ratios of PUI densities obtained using these two models. Since, however, the adopted solar wind speeds were identical, the PUI density ratios are equivalent to the PUI flux ratios.

These ratios are shown in Figures~\ref{fig:density1}--\ref{fig:density3} with dashed lines. Generally, the pattern of the PUI density and flux differences follows the pattern of ISN H density differences, but because the PUI density at a given location in space $\vec{r}$ is due to an integral over the radial line from the Sun, and most of the differences between the two models are at close solar distances, then the spatial extent of the differences between the two models is a little larger, e.g., while at 1~AU downwind the IKL18/ST09 ISN H density difference in 1996.0 was less than 10\%, the equivalent PUI density difference is $\sim 15$\%. 

\section{Consequences for selected aspects of heliospheric studies}

With the differences between the ISN H and PUI densities inside the heliosphere resulting from using the IKL18 and ST09 models, we verify if conclusions from selected earlier studies, drawn based on modeling of ISN H carried out using the ST09 model, need revision. 

\subsection{The density of ISN H at the termination shock}

One of the very important results obtained using the ST09 radiation pressure model is the estimate of ISN H density at the termination shock based on Ulysses pickup ion observations \citep{bzowski_etal:08a}. This quantity had been obtained by fitting the hydrogen PUI observations performed by SWICS/Ulysses in an orbital arc between the solar pole and the ecliptic plane. \citet{bzowski_etal:08a} argued that this section of the Ulysses orbit was close to the ISN cavity boundary, where the ISN H density is close to $\exp(-1)$ of its value at the termination shock. In this geometric location, the observed production rate of PUIs, obtained directly from the measured distribution function of PUIs, is proportional to the density of ISN H at the termination shock, and the proportionality coefficient is the ionization rate of ISN H at the location of the measurement, which is known. 

\citet{bzowski_etal:08a} argued that their density determination is  weakly sensitive to potentially unknown details of the solar radiation pressure and ionization rate. However, since the time of their measurement (1) the model of the heliospheric ionization rate has been upgraded (from that in \citet{bzowski_etal:08a} to that in \citet{sokol_etal:13a}), (2) the radiation pressure model has changed (from ST09 to IKL18), and (3) the Ulysses orbit was in the region sensitive to the change in radiation pressure model, as suggested by the green and magenta lines in Figure \ref{fig:density2}, we repeated the calculations presented by \citet{bzowski_etal:08a}. We found that the ISN H density at the termination shock derived from the original observations and our present calculations is $n_{TS}=0.094 \pm 0.023$ cm$^{-3}$ and it differs from that obtained by \citet{bzowski_etal:08a} by  less than 10\%. Since the density determination uncertainty is at the level of $\sim 25$\%, mostly based on the uncertainty of the PUI measurements, those two numbers are statistically consistent. We have no grounds to claim our present determination is more accurate than the original one.

\subsection{The vector flux of ISN H along the Earth's orbit}
\begin{figure}
\centering
\includegraphics[width=0.85\columnwidth]{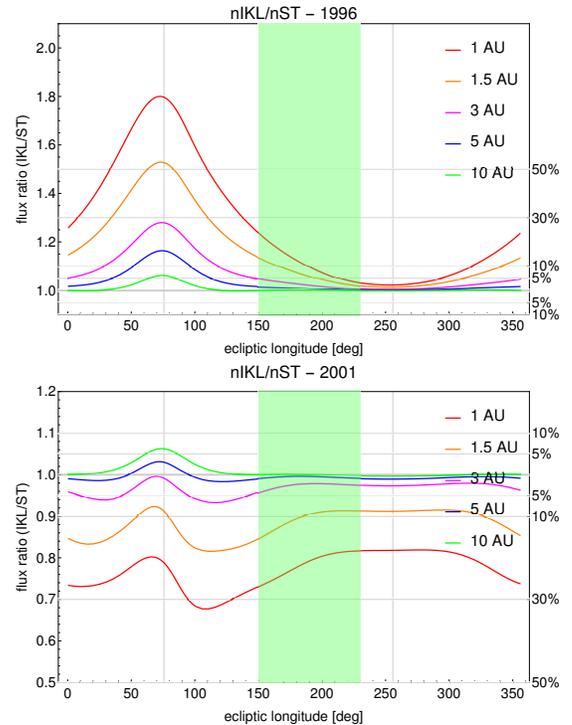}
\caption{
%{\em{fig:flux1}} 
The ratio of ISN H and D fluxes calculated using two radiation pressure models (IKL18 and ST09) as a function of ecliptic longitude. Results are shown for different distances from the Sun. The black line marks the flux ratio for deuterium at 1~AU, while hydrogen is represented by the color lines: red line is corresponding to 1~AU, orange, to 1.5~AU, magenta to 3~AU, blue to~5 AU, and finally green line represents 10~AU. Top panel shows the solar minimum conditions (1996), while bottom panel presents the solar maximum conditions (2001). Green shadow marks the IBEX-Lo observation range in the ecliptic plane.
}
\label{fig:flux1}
\end{figure}

The flux vector of the neutral gas is calculated as follows:
\begin{equation}
\vec{F}=\vec{v}_{\text{pr}}\,n_{\text{pr}}+\vec{v}_{\text{sc}}\,n_{\text{sc}},
\end{equation}
where $\vec{v}_{\text{pr}}$ and $\vec{v}_{\text{sc}}$ are the velocity vectors of the primary and secondary populations, and $n_{\text{pr}}$ and $n_{\text{sc}}$ are their densities. Analysis of the IKL18/ST9 ratio of this quantity along the Earth's orbit facilitates estimating how much the model signal corresponding to the ISN H flux directly sampled by IBEX is sensitive to the change of radiation pressure model. We show this ratio in Figure~\ref{fig:flux1} for several distances in the ecliptic plane.

The shape and character of changes visible on Figure \ref{fig:flux1} is similar to those for the density, shown in Figure \ref{fig:density1}. This is not surprising since flux is a product of density and speed, and the latter one is not changing significantly (see Section~\ref{sec:speed}). The flux differences along the Earth orbit are larger than 10\% in a large portion of the Earth's orbit, which likely affects the quantitative interpretation of direct-sampling observations. We marked the area where IBEX collects data by the light green shading in Figure~\ref{fig:flux1}. The differences between the two models in this region are not negligible, both for the low and high solar activity conditions. Therefore, in Section~\ref{sec:IBEX} we investigate the flux of atoms impacting the IBEX-Lo detector in greater detail.

\subsection{Deuterium}
The abundance of deuterium relative to hydrogen is important for Big Bang models and for studies on the chemical evolution of interstellar matter near the solar Galactic neighborhood. \citet{tarnopolski_bzowski:08a} and \citet{kubiak_etal:13a} investigated the expected flux of ISN D at 1~AU and the observability of this population by IBEX-Lo. Based on these estimates, \citet{rodriguez_etal:13a, rodriguez_etal:14a} after challenging, meticulous analysis claimed that out of several D atoms found in the IBEX data, approximately three can be interpreted as interstellar. Even though this statistics is very low, we verify if this claim holds based on the insight obtained from the IKL18 more accurate model of radiation pressure. 

We calculated the prediction for the flux of ISN D along the Earth's orbit using the IKL18 and ST09 models and the present model of ionization. The ratio of the fluxes is shown by the black line in Figure~\ref{fig:flux1}. While some differences are visible in this region, they most likely average out to zero over the region of IBEX observations and hence the ISN D detection claim by \citet{rodriguez_etal:13a} is supported. 

\subsection{The flux of ISN H at IBEX}
\label{sec:IBEX}
\begin{figure*}
\centering
\includegraphics[width=0.9\textwidth]{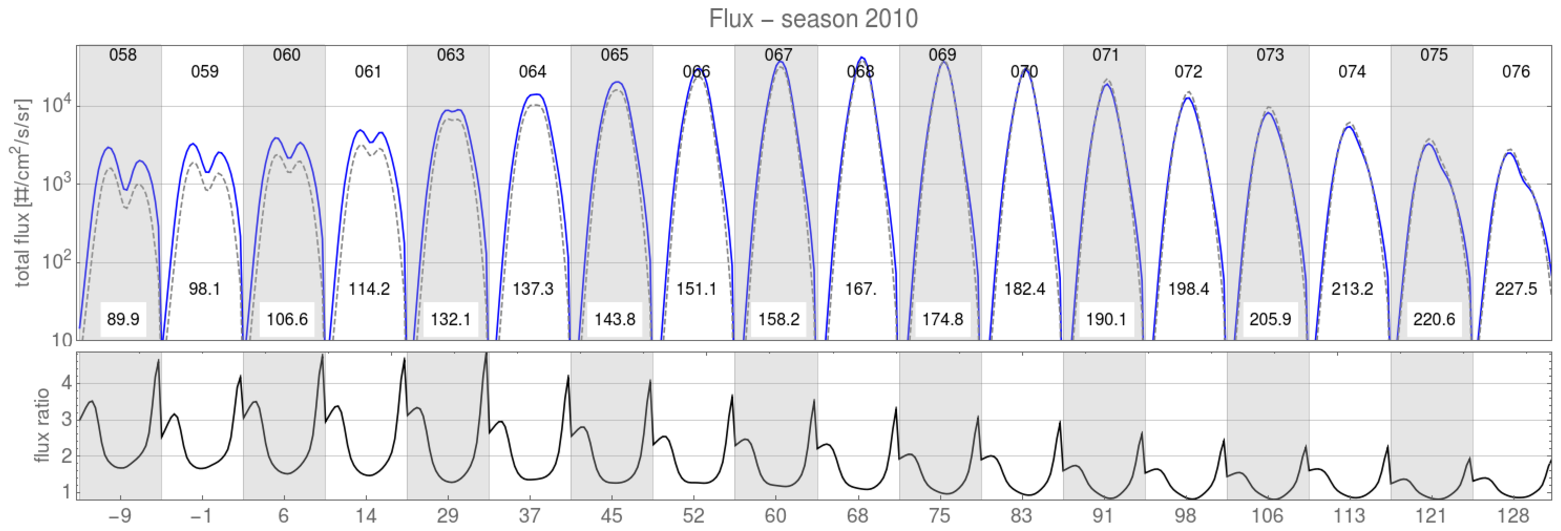}
\includegraphics[width=0.9\textwidth]{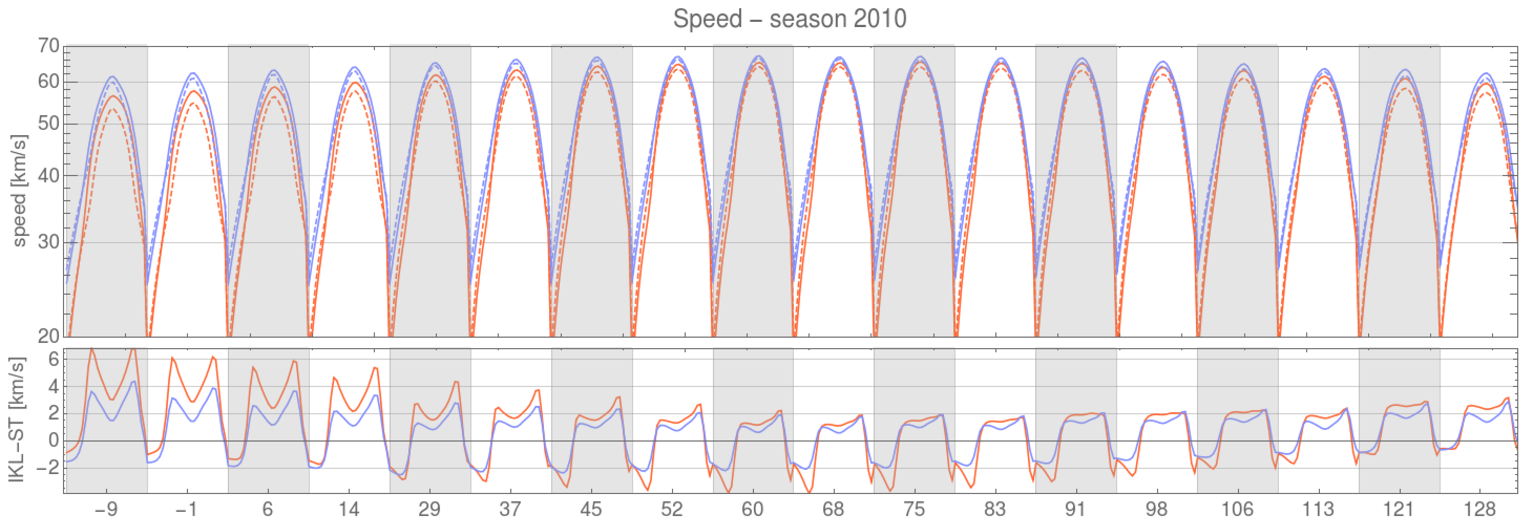}
\includegraphics[width=0.9\textwidth]{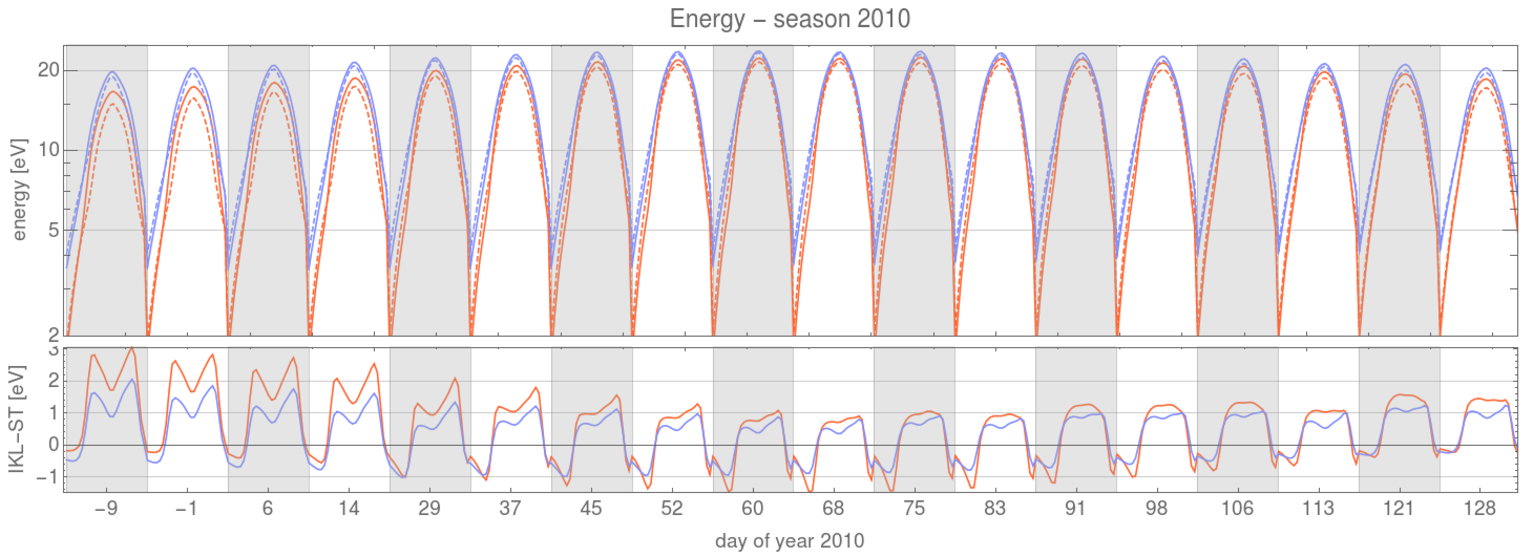}
\caption{
%{\em{fig:ibexFlux2010}}
The total flux (a sum of the fluxes of the primary and secondary populations, upper panel), mean speed (separately for the primary and secondary populations, middle panel) and mean energy (separately for the primary and secondary populations, lower panel) of ISN H filtered by the collimator of IBEX-Lo. The quantities shown were simulated for each 6-degree spin angle interval within the range $180\degr-330\degr$, increasing from left to right within each vertical strip for IBEX ISN observation season 2010 (low solar activity conditions) using the ST09 and IKL18 models of radiation pressure. The blue solid line in the upper panel corresponds to the IKL18 model and gray dashed line presents the simulations for the ST09 model. In the middle and lower panels, the red color represents the primary population and blue color the secondary population. Solid lines correspond to the IKL18 model, and dashed lines to ST09. The simulations representing individual orbits are marked by the white and gray strips. The spin angle ranges are identical for all presented orbits. The orbit numbers are shown at the top of the first panel, and the mean ecliptic longitudes of the Earth for individual orbits are presented at the bottom of the first panel. The lower subpanel in the first panel represents the ratio of IKL18 to ST09 model fluxes, and the lower subpanels in the second and third panels show the differences in speed and energy, respectively, predicted by the IKL18 and ST09 models. The simulations for individual orbits are arranged chronologically during the season.
}
\label{fig:ibexFlux2010}
\end{figure*}

\begin{figure*}
\centering
\includegraphics[width=0.9\textwidth]{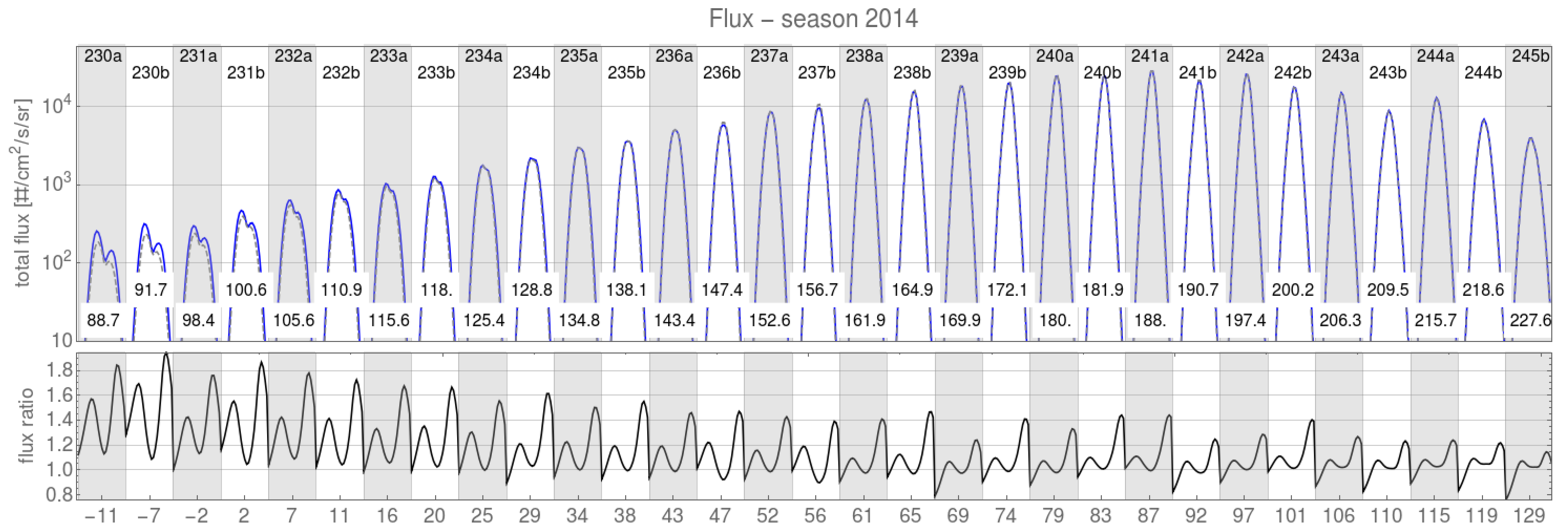}
\includegraphics[width=0.9\textwidth]{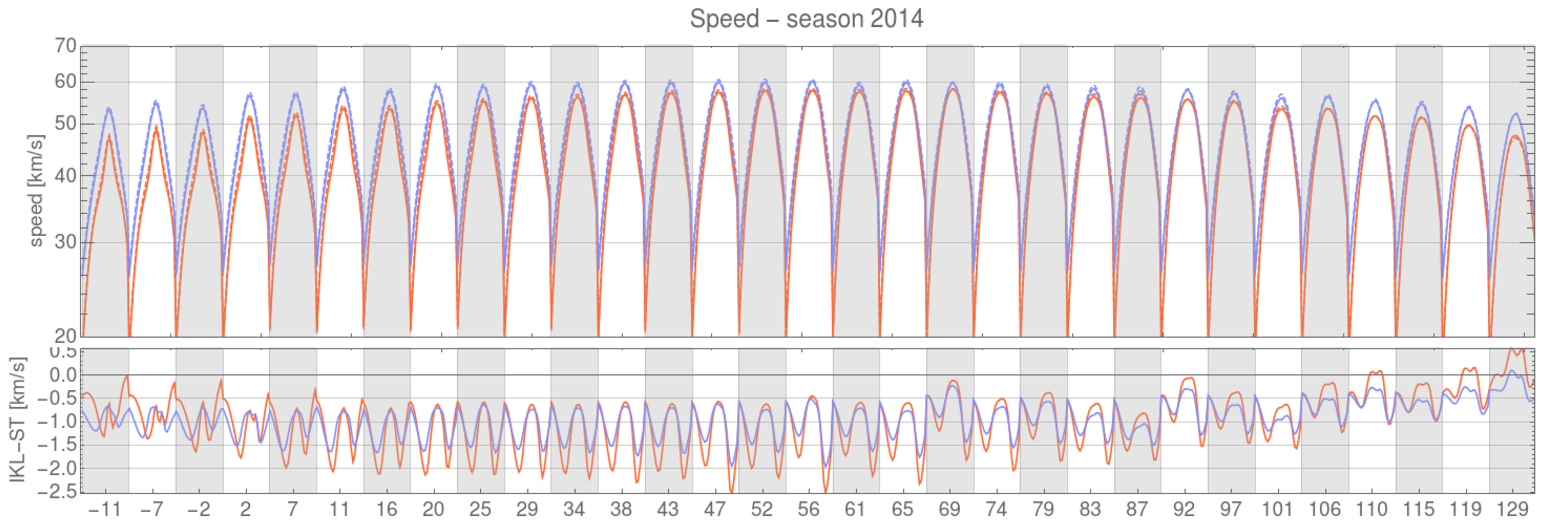}
\includegraphics[width=0.9\textwidth]{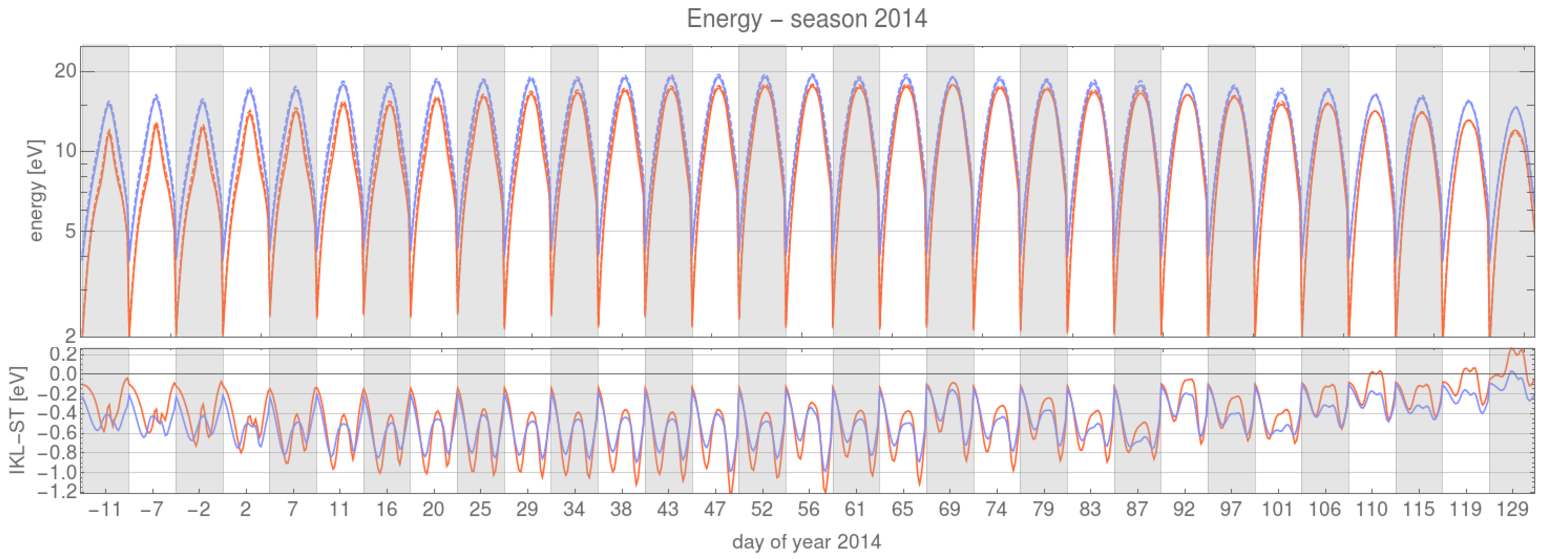}
\caption{
%{\em{fig:ibexFlux2014}}
Similar as Figure~\ref{fig:ibexFlux2010}, but for the ISN observation season 2014. Note that in 2014, the observations in each IBEX orbit were split between two orbital arcs, marked a and b.
}
\label{fig:ibexFlux2014}
\end{figure*}

In this section, we analyze the quantities relevant for direct-sampling observations of ISN H by IBEX. IBEX \citep{mccomas_etal:09a} is the first satellite to directly sample ISN H in the Earth's orbit. IBEX is a spin-stabilized spacecraft with the spin axis changed once (until 2012) or twice per orbit to approximately follow the Sun \citep{mccomas_etal:11a}. The ISN atoms are observed using the IBEX-Lo instrument \citep{fuselier_etal:09b}, which is a time of flight mass spectrometer. Before entering the detector, the observed atoms pass the collimator, which defines the field of view of the instrument. The data are collected while the spacecraft is rotating, and the observed counts are binned into time-intervals with the length selected so that they correspond to fixed spin angle bins. Since between the repositionings the spin axis is fixed in space, the observations during an individual orbital arc cover a specific, fixed region in the sky. 

Since the spacecraft is traveling with the Earth around the Sun, the relative velocities of the atoms entering the IBEX-Lo detector within an individual spin angle bin are vector sums of the individual atom velocity and the velocity of the spacecraft relative to the Sun. As a result, the mean speeds of atoms observed in individual spin angle bins in a given orbit differ from each other. Since the ISN H gas features a thermal spread, the speeds of the ISN H atoms entering the IBEX-Lo detector within a given spin angle bin feature a certain spread, which varies from orbit to orbit and as a function of the spin angle. Due to this thermal spread, the mean energies of the atoms within a spin angle bin are a little larger than the energies calculated from the mean speed of these atoms.  

The sensitivity of IBEX-Lo decreases with the decrease of the energy of impacting atoms and therefore the observations are possible only in a portion of the Earth's orbit where the orbital velocity of the spacecraft and the Earth adds to the flow speed of ISN H. The region of ecliptic longitudes where these observations are possible is indicated by the green shadow in Figure~\ref{fig:flux1}. 

The issue of the influence of radiation pressure on ISN H atoms is crucial for the interpretation of IBEX observations of ISN H, as suggested by \citet{schwadron_etal:13a} and \citet{katushkina_etal:15b}. These authors found that there is a very large discrepancy between the proportions of the ISN H signal found in IBEX-Lo energy steps 1 and 2 observed during the 2009 and 2010 ISN seasons on one hand and the simulated fluxes for these energy steps, obtained using a state of the art model of ISN H distribution \citep{izmodenov_alexashov:15a}. These differences can be partly, but not fully mitigated when one assumes a larger horn-to-reversal ratio in the solar Lyman-$\alpha$ line profile. In their analysis, \citet{schwadron_etal:13a} and \citet{katushkina_etal:15b} used the line profile from the ST09 model and adopted a very similar parameters for the primary and secondary ISN H populations and the termination shock to those that we use in this paper. Here, we investigate how the employment of the IKL18 model changes the predicted flux of ISN H entering the IBEX-Lo detector.

In the simulations carried out using the nWTPM code \citep{sokol_etal:15b}, the distribution function of ISN H atoms just before entering the detector is represented by a superposition of distribution functions corresponding to the primary and secondary populations of ISN H, calculated for the time of detection and the location and velocity of the detector in space. In the presented simulations, we present the total flux of ISN H entering the detector, but we show the mean speeds and energies of the two populations separately, because this additional information may be of significance for the understanding the IBEX observations. 

Even though the ISN observations are carried out during yearly seasons, the observed signal features large variations between the seasons. This is on one hand because of physical reasons \citep[the ISN H is heavily modulated in the Earth's orbit because of the variations in the ionization rate and radiation pressure,][]{rucinski_bzowski:95b, bzowski_etal:97, tarnopolski_bzowski:09, bzowski_etal:13a}. On the other hand, because the observation conditions never repeat precisely from one year to another, the ISN H signal observed is very sensitive to small differences in the spin angle pointing and the length of ``good times'' during the observations. 

Since the sensitivity of the IBEX-Lo detector to H atoms is a strong function of the atom impact energy, the atom energy must be appropriately taken into account when simulating the observed counting rates. The observations are carried out with the instrument sequentially switched between eight energy channels. The ISN signal was observed in the four lowest-energy channels \citep{galli_etal:15a}. The signal in the instrument is due to H$^-$ ions leaving the specially prepared conversion surface and entering the electrostatic analyser. These ions originate either from direct conversion of neutral H atoms hitting the conversion surface, or are being sputtered by He atoms. It is not possible to differentiate between the ions from these two sources without meticulous statistical analysis of times of flight of the observed H$^-$ ions \citep{rodriguez_etal:13a, rodriguez_etal:14a, park_etal:15a}. Therefore, ISN H can only be clearly identified during a portion of the ISN sampling season, when the spacecraft is in the region of Earth's orbit where the ISN He flux is very weak \citep{saul_etal:12a}. This region begins at ecliptic longitude $\sim 175\degr$. The yearly peak of the ISN H signal is obscured by the signal from ISN He, and the signal in the slope of ISN H is particularly sensitive to details of radiation pressure, as we show further in this paper. 

Because of these complexities, we investigated the ISN H gas entering the IBEX-Lo detector in greater detail. In the simulations, we obtained the atom flux filtered by the collimator but before its further processing inside the instrument. The simulations have been carried out for the conditions of low (2010) and high (2014) solar activity, separately for the primary and secondary ISN H populations. Results are shown in Figure~\ref{fig:ibexFlux2010} for 2010 and Figure~\ref{fig:ibexFlux2014} for 2014. 

Already from analysis of differences between the predictions of ST09 and IKL18 for the total flux at 1~AU in the region covered by IBEX (Figure~\ref{fig:flux1}, the green shading) one can expect that the simulated IBEX signal will be sensitive to the change from one radiation pressure model to the other. And sure enough, the differences in the simulated flux for both seasons are considerable. They are strong functions of the spin angle on one hand, but also of the location along the Earth's orbit on the other hand. These differences are visible both for low and high solar activity, as illustrated in the top panels of Figures~\ref{fig:ibexFlux2010} and \ref{fig:ibexFlux2014}. 

Before proceeding to further discussion we point out some relevant findings from analysis of IBEX observations. First, we remind of the discrepancies between model and observations in different energy steps pointed out by \citet{schwadron_etal:13a} and \citet{katushkina_etal:15b}. Second, \citet{galli_etal:17a} found a considerable amount of ISN H in the ISN data from 2009 and 2010 seasons, but practically no ISN H at all in the data from 2014. Third, \citet{kubiak_etal:14a, kubiak_etal:16a} analyzed the IBEX signal from the portion of the Earth orbit where the Warm Breeze prevails. The Warm Breeze is the secondary population of ISN He \citep{kubiak_etal:16a, bzowski_etal:17a}, but the data vs model comparison by \citet{kubiak_etal:16a} showed a considerable residual, especially during the low solar activity observation seasons. Fourth, \citet{park_etal:16a} found, based on a detailed statistical analysis of the IBEX-Lo signal, that there was a certain contribution from genuine H atoms to the Warm Breeze signal during the low-activity seasons.

Careful inspection of Figures~\ref{fig:ibexFlux2010} and \ref{fig:ibexFlux2014} brings the following conclusions. The flux obtained using the IKL18 model is larger for low solar activity (2010) than the flux obtained using ST09. The differences vary with the spin angle and location along the Earth orbit, but they are considerable everywhere. These differences persist also during the 2014 season of high solar activity, but for this season the flux ratio is lower. The largest differences in the magnitude of the flux persist in the Earth orbit portion where the secondary population of ISN He (the Warm Breeze) is observed (orbits 58-66 of season 2010 and orbits 230-236 of season 2014), with the IKL model predicting a much larger flux in this region than TB09 does. 

The differences between the energies of the ISN H atoms entering the detector between IKL18 and ST09 are the largest during the 2010 season, especially in the Warm Breeze portion of the Earth's orbit, where they are on the order of 1--2~eV, for the total energy of the impacting atoms varying from 5 to almost 20~eV. They drop approximately by half in the portion of Earth's orbit where ISN H was observed without obscuration from ISN He (orbits 70-76 of season 2010 and orbits 239-245 for season 2014). Generally, the energies of ISN H atoms entering the detector are comparable between this latter portion of the Earth's orbit and the WB portion. This suggests that since ISN H is visible late during the season, then it must be present in the signal also in the WB part of the orbit, when and where the energies and flux magnitudes are similar.

The magnitude of the flux entering the detector at the ISN H portion of the orbit during 2014 is similar to that during 2010, or even larger. This conclusion agrees between IKL18 and TB09. However, the ISN H signal has not been observed during 2014. The explanation may be the strong increase in the sensitivity of IBEX-Lo to ISN H atoms with increasing energy. The energy of ISN H impacting the detector in 2014 is less by $\sim 4$~eV than the energy in 2010 ($\sim 16$~eV vs $\sim 20$~eV at peak). The differences between the two models in the predicted energies are on the order of 1--2~eV in 2010 (IKL18~$>$~ST09) and drop to $\sim -0.2$~eV in 2014 (IKL18~$<$~TB09). So the differences in the energy between the two radiation pressure models are less than the differences in the energy between 2010 and 2014. Since the expected fluxes are similar for the two seasons, we conclude that the reason why ISN H has not been detected in 2014 may be an abrupt drop in the sensitivity of IBEX-Lo for the H atom energies below $\sim 20$~eV, and additionally a two-fold drop in the IBEX-Lo sensitivity after 2012 because of the reduction in the post-acceleration voltage. This latter sensitivity reduction was demonstrated by \citet{swaczyna_etal:18a} for ISN He observations.

Both IKL18 and ST09 predict a large drop in the ISN H flux and a little lower drop in the ISN H energy during 2014 in the WB region of the Earth's orbit. Therefore, the ISN signal observed in this region during high solar activity is very likely free from contribution from ISN H and therefore is favorable for analysis of the secondary population of ISN He atoms. 

Based on this insight we speculate that the change in the simulations to a more refined model by IKL18 will not be sufficient to resolve the data/simulation discrepancy reported by \citet{schwadron_etal:13a, katushkina_etal:15b}. Indeed, as these authors suggest, a larger horn/minimum ratio may be needed in the radiation pressure model. Since the analysis of the solar line profile observations by IKL18 does not allow this ratio to be sufficiently high, in the next section we suggest an effect, up to now largely neglected, that might help alleviate the data/simulation discrepancy. 

\section{Absorption of the Lyman-$\alpha$ spectral flux by ISN H and the resulting change of effective radiation pressure with the solar distance}
\label{sec:absorption}

Due to the presence of ISN H in the heliosphere, the solar spectral flux in the Lyman-$\alpha$ line is differently absorbed in different frequencies, depending on the prevailing radial velocities of ISN H atoms  \citep{wu_judge:79b,hall:92a,brasken_kyrola:98,quemerais:00}. The magnitude of this effect is correlated with the distribution of the density and radial velocity of the gas in space. We calculated the differential absorption of the solar spectral flux following the procedure given by \citet{quemerais:06a}. To that end, we computed the local distribution function of ISN H (the primary and secondary populations) in each point of our calculation grid. Then, in each grid point we have projected the three-dimensional distribution function on the radial direction and assumed that this new one-dimensional projected function is consistent with the normal distribution with a certain thermal spread and radial velocity, different between the grid points. 

The absorption for a point at a certain distance from the Sun in a given direction is a superposition of the absorption contributions from all portions of ISN H between the Sun and the chosen distance, integrated over the radial line, separately for all radial speeds. As a result of this differential absorption, a characteristic absorption feature appears in the solar Lyman-$\alpha$ line, with the depth increasing with the column density of ISN H between the Sun and the given point in space. The effective profile of the solar Lyman-$\alpha$ line at a location given by the location $r, \Omega$ in space is given by Equation~\ref{eq:abs}
\begin{eqnarray}
\label{eq:abs}
I(r,\Omega,\nu)&=&I_0(\Omega,\nu) \exp\left(-\int_{R_{\odot}}^r n(r',\Omega) \sigma(\nu)dr'\right), \\
\sigma(\nu)&=&\sigma_0 \exp\left(-\left(\frac{\nu -\nu_r(v_r)}{\Delta \nu_D}\right)^2\right),\\
\Delta\nu_D&=&\frac{\nu_0}{c}\sqrt{\frac{2kT_g}{m}}.
\end{eqnarray}
$I(r,\Omega,\nu)$ is the flux of photons of frequency $\nu$ that are present at the distance $r$ in the direction $\Omega$, $I_0(\Omega,\nu)$ is the initial flux emitted from Sun's surface, $n(r',\Omega)$ is the density of ISN H at the considered point, $\sigma(\nu)$ is the cross-section for the absorption of a photon of frequency $\nu$ by hydrogen atom of radial velocity $v_r$, $T_g$ is the temperature of the gas. 

The width of the absorption feature depends on the thermal spread and the radial velocity of ISN H, which both vary with the distance from the Sun and with the angle off the upwind direction. Therefore, the absorption features have different center wavelengths and different widths. The absorbed photons are redistributed in direction and frequency, thus forming the heliospheric resonant backscatter glow. This topic has been addressed in several papers \citep[e.g.,][]{wu_judge:79b,hall:92a,quemerais_bertaux:93a, brasken_kyrola:98, quemerais:00, scherer_fahr:96, fayock_etal:15a}, but it is outside the scope of this paper.

The wavelength-selective absorption of the solar spectral flux results in a reduction in the radiation pressure that ISN H atoms effectively sense. Since the absorption is by atoms that make the bulk of ISN H, the effective radiation pressure force acting on a typical ISN H atom will be reduced more than the purely-geometric reduction of the solar Lyman-$\alpha$ flux by $1/r^2$. However, the $1/r^2$ reduction is the basis of the concept that the radiation pressure force effectively compensates the solar gravity force identically for all distances and that this reduction is only a function of time (because of the varying intensity of solar radiation) and radial speed (because of the Doppler effect ``shifting'' the atoms along the Lyman-$\alpha$ profile). With the actual absorption taken into account, the effective radiation pressure force falls off with the solar distance more rapidly than the $1/r^2$ dependence typically used up to now. 

\begin{figure*}
\centering
\includegraphics[width=0.9\textwidth]{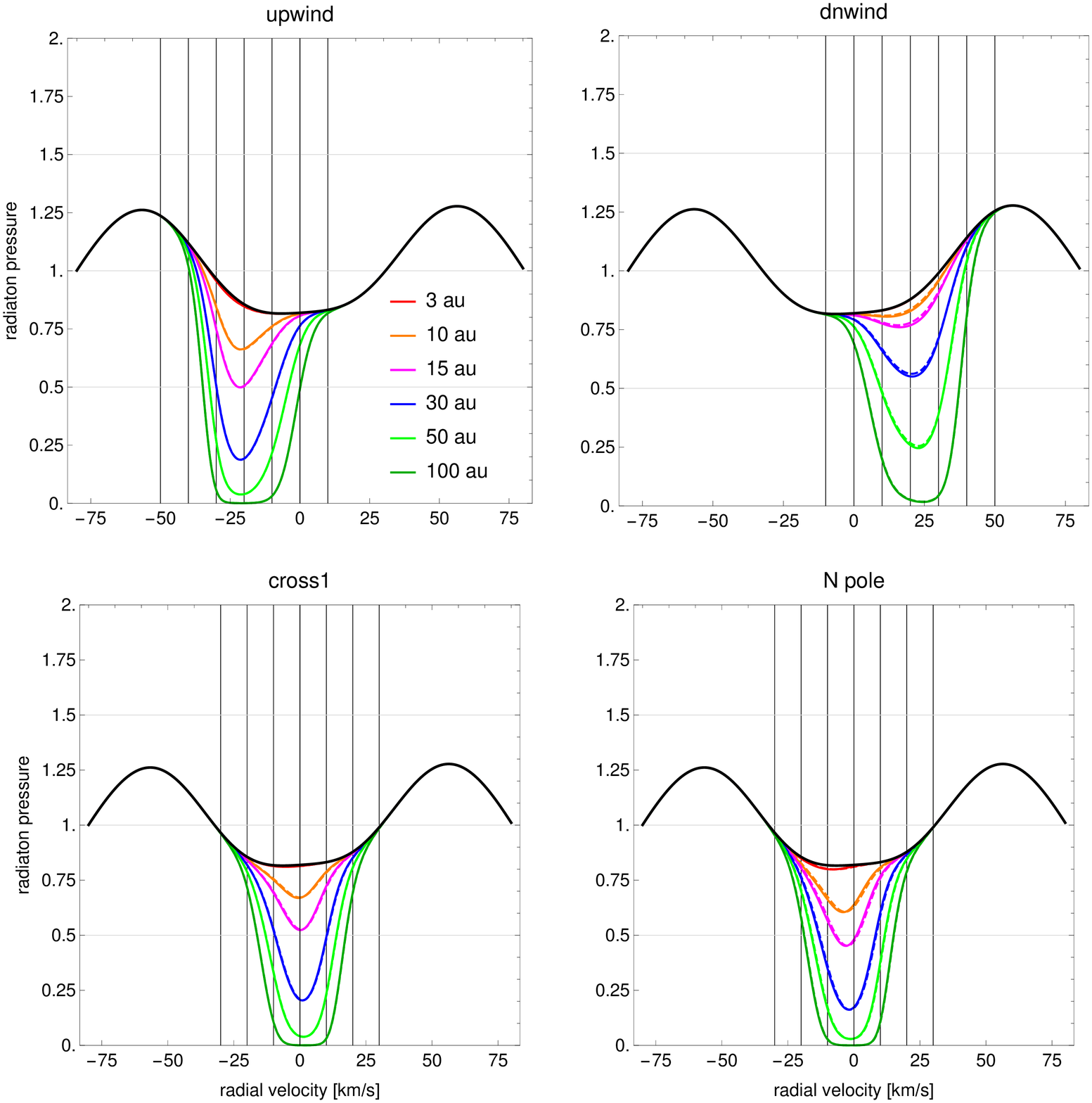}
\caption{%{\em{fig:absProfile1996}}
Radiation pressure profiles for selected distances from the Sun, modified by the absorption by ISN H in the heliosphere, simulated for the solar minimum conditions (1996). The four panels present the absorption for the four selected directions in space, defined in Table~\ref{tab:directions}: ``upwind'' -- top left, ``dnwind'' -- top right, ``cross1'' -- bottom left, and ``Npole'' -- bottom right. The solid black line represents the original profile observed at 1~AU, the other colors are described in the top left panel. The solid lines represent the IKL18 model, the dashed lines the ST09 model. Vertical grid lines sample central part of the profile line every 10 km s$^{-1}$. The adopted density on termination shock (90AU) is n$_{TS}=0.0851$ cm$^{-3}$.
}
\label{fig:absProfile1996}
\end{figure*}

\begin{figure*}
\centering
\includegraphics[width=0.9\textwidth]{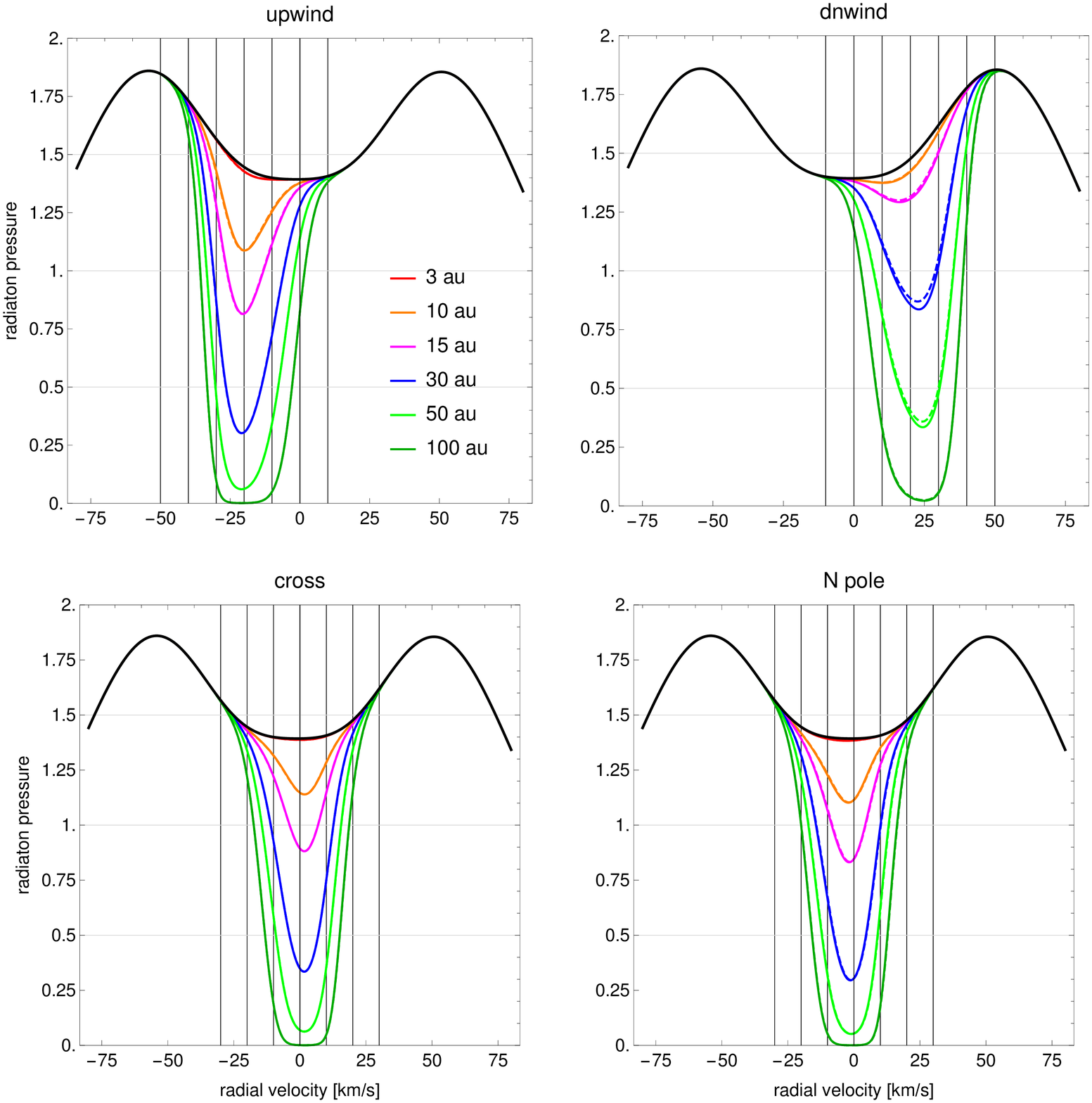}
\caption{%{\em{fig:absProfile2001}}
As Figure~\ref{fig:absProfile1996} but for the solar maximum conditions (2001).
}
\label{fig:absProfile2001}
\end{figure*}

To investigate the absorption, we adopted the four directions related to the geometry of the inflow, defined in Table~\ref{tab:directions}. The absorbed line profiles for these directions are presented in Figure~\ref{fig:absProfile1996} in the radiation pressure units for the solar minimum, and in Figure~\ref{fig:absProfile2001} for the solar maximum conditions. Each panel presents absorption profiles for one of these antisolar lines, color-coded for a number of solar distances. Solid lines correspond to the IKL18 model, and dashed lines to ST09. 

The simulations suggest (see Figures~\ref{fig:absProfile1996} and \ref{fig:absProfile2001}) that for distances below $\sim 3$~AU, the solar Lyman-$\alpha$ line is unchanged (i.e., the absorption is insignificant), because there is not enough gas to absorb a significant fraction of the photons. With increasing distance from the Sun, the absorbed part of the spectrum becomes deeper and wider because there is more gas, which has different radial velocities. Depending on the direction in space, the wavelength of the absorption centroid varies following the effective radial speed of the gas for this direction. At the upwind direction, the absorption is blue-shifted because the gas is approaching the Sun, at the crosswind directions the shift is almost null because the radial speed of the gas just passing the Sun is close to zero, and at the downwind direction the absorption is red-shifted because the gas is flowing away from the Sun. Note a small but well visible difference in the centroid of the ``cross1'' and ``npole'' absorption features : they are caused by the stronger ionization close to the ecliptic plane than at polar latitudes. As discussed in \citet{bzowski_etal:97}, stronger ionization losses modify the bulk speed and flow direction of the gas, masquerading for a dynamical effect.

\begin{figure*}
\centering 
\includegraphics[width=0.95\textwidth]{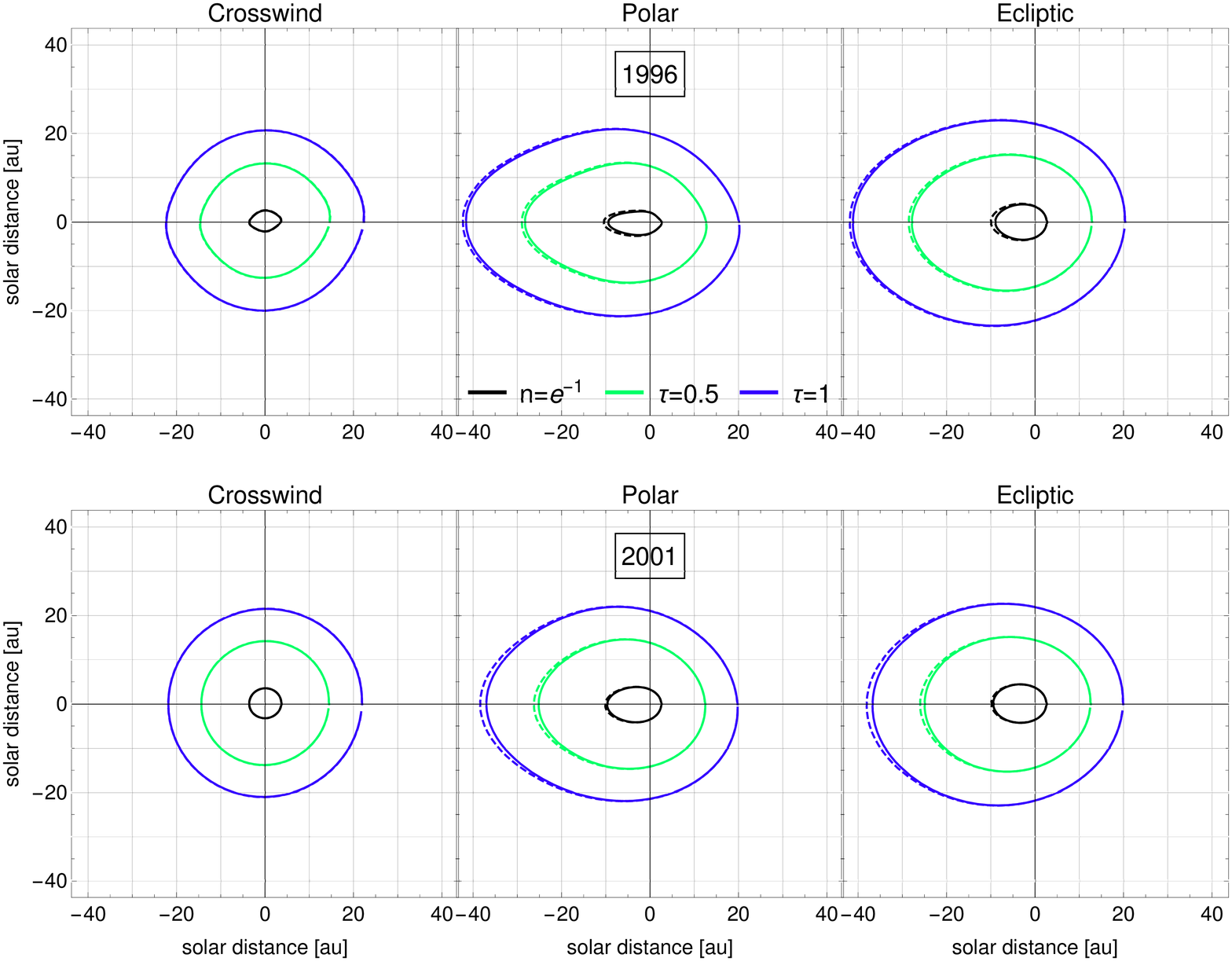}
\caption{
%{\em{fig:contour} }
Isocontours of the optical depth of the absorption for the optical depth $\tau=0.5$ (green) and $\tau=1$ (blue), presented in the crosswind plane in the left column, polar plane in the middle column, and the ecliptic plane in the right column. Top row corresponds to the solar minimum conditions (1996), the bottom row to the solar maximum conditions (2001). Also shown are the contours of the ISN H cavity in the three planes, i.e., the geometric locations where the local density of ISN H is reduced to $1/\mathrm{e}$ of its value at the nose of the termination shock (black lines). The solid lines represent the IKL18 model and the dashed lines the ST09 model.}
\label{fig:contour}
\end{figure*}

With this insight, a compelling question appears where in space this absorption becomes significant. Figure~\ref{fig:contour} shows isocontours of the absorption expressed as the optical depth for three representative planes in the heliosphere: the ecliptic plane, the polar plane, and the crosswind plane (defined as the plane containing the Sun and perpendicular to the upwind-downwind vector). The green contour corresponds to optical depth $\tau = 0.5$, i.e., the region where $\sim 60$\% of photons are transmitted ($I=I_0 \mathrm{e}^{-\tau}$, where $I_0$ is the photon flux measured at 1~AU), and the blue contour to $\tau = 1$, conventionally considered as the optical depth corresponding to optically thick gas, where $\sim 36$\% of photons are transmitted (i.e., the radiation pressure is reduced to 36\% of its value at 1~AU). These regions are much larger in all three planes considered than the boundary of ISN H cavity, defined as the geometric location where the ISN H density is less than $1/{\mathrm{e}}$ of its magnitude at the termination shock.

This simulation suggests that ISN H is optically thin within a region much larger than the size of the cavity. The size of the cavity region and of the intermediate and large optical thickness of the gas evolve a little during the solar activity (especially the cavity). IKL18 and TB09 give very similar predictions for these regions, with some differences visible mostly in the downwind region of the heliosphere. In these large distances from the Sun the role of radiation pressure for the ISN H distribution is mild. However, earlier in this paper we have shown that the distribution of ISN H inside the heliosphere is sensitive to much finer details of radiation pressure than these related to the absorption at $\tau = 0.5$. 

This suggests that neither the absorption simulations nor the ISN density simulations that we have made are self-consistent. We simulated the absorption based on a model of ISN H parameters calculated assuming that the entire heliosphere is optically thin, which appears to not be the case in a large portion of the heliosphere. Therefore the conclusions we offer are tentative. Providing a definite answer will be possible when a simulation with an appropriately modified radial profile of radiation pressure has been done. This, however, is now left for a future study. 

Nevertheless, the insight we have obtained is encouraging. It seems that the effective radiation pressure may be indeed different than thought up to now. The changes are likely to modify the horn-to-minimum ratio, as postulated by \citet{schwadron_etal:13a} and \citet{katushkina_etal:15a} to explain the IBEX direct-sampling observations of ISN H). So far it was assumed that the horn height is larger than observed by \citet{lemaire_etal:15a}. We suggest a different hypothesis, namely the absorption on ISN H reduces the minimum value of the effective solar line profile perceived by ISN H atoms. It seems that details of the solar Lyman-$\alpha$ line profile (IKL18 vs ST09) will be of secondary importance here because our simulations suggest that the magnitude of absorption is little sensitive to these details. This is because most of the absorption occurs in the regions where differences in the ISN H distribution due to IKL18 vs ST09, discussed in Section~\ref{sec:gasParams}, become insignificant. 

%\begin{figure*}
%\centering
%\includegraphics[width=0.6\textwidth]{figures/figAbs.eps}
%\caption{{\em{fig:Abs} } 
%}
%\label{fig:Abs}
%\end{figure*}

\section{Summary and conclusions}
We found that ISN H is sensitive to the seemingly small differences in radiation pressure between the IKL18 and ST09 models. Using the new model of radiation pressure developed by \citet{kowalska-leszczynska_etal:18a} has significant consequences for the densities and related parameters of ISN H and PUIs in selected regions of space. The most affected is the downwind region, where the IKL18/ST09 ratio of predicted densities of ISN H in the Earth's orbit can be as large as 2 during low solar activity phase, and the region where the differences are at least 10\% extends to $\sim 7$~AU. The behavior of the model ISN H flux differences follows the behavior of the density differences, and PUIs are affected even a little farther away from the Sun. The differences between the model predictions exist for all phases of solar activity. On the other hand, differences between the two models for the bulk velocity and its components are relatively small, on the order of $\sim 10$\%, or just a few km~s$^{-1}$. 

Despite the region of significant differences between ISN H and PUI density includes the Ulysses orbit, the magnitude of ISN H density at the termination shock, obtained by \citet{bzowski_etal:08a} based on analysis of PUI observations at Ulysses using ISN H models with the ST09 radiation pressure model does need not revision because the magnitude of ISN H density at the termination shock we have derived now using the IKL18 model of radiation pressure is safely inside the uncertainty range reported by \citet{bzowski_etal:08a}.

The simulated fluxes of ISN H atoms hitting the IBEX-Lo detector are noticeably different when calculated using the IKL18 model, particularly during the early portions of the yearly ISN observation seasons, when the secondary population of ISN He (the Warm Breeze) is observed. The IKL18 model predicts a considerably larger H flux in these regions during low solar activity. The simulations of the flux at IBEX carried out using the IKL18 radiation pressure model suggest that the IBEX signal observed during the first part of the 2010 season (orbits 58-66) consists of helium Warm Breeze and hydrogen, while in season 2014 (orbits 230-236) there is only the helium component. Therefore IBEX observations of the Warm Breeze from the seasons of high solar activity are favorable for analysis of the secondary component of ISN He since they seem to be free from contamination by ISN H. 

By comparing the simulated fluxes and energies of IBEX H atoms between the low- and high solar activity seasons in the portions  of the Earth orbit where ISN H is observed we found that there should be a threshold in the energy sensitivity of IBEX-Lo somewhere below $\sim 20$~eV. This is because in 2014, when ISN H was not detected, the simulated energy of ISN H in this portion of the orbit is lower than 20~eV and lower than the $\sim 20$~eV energy of ISN H in 2010. This could explain why in 2010 ISN H was detected, while in 2014 it was not despite a very similar level of the simulated ISN H flux. 

Differences between the two considered models of radiation pressure are not negligible, but they may become less important than the effects of absorption of the Lyman-$\alpha$ solar line by the ISN H in the heliosphere. We have studied the influence of this absorption on radiation pressure acting on ISN H atoms in the heliosphere. The importance of the absorption increases with distance from the Sun. The environment become optically thick (optical depth $\sim 1$) at more than $\sim 20$~AU from the Sun. The ISN H cavity and all the effects connected with differences between radiation pressure models are within optically thin environment.  

Therefore we finally conclude that radiation pressure acting on ISN H in the heliosphere is not understood as well as it has been thought. Future studies of ISN H inside the ISN H cavity must take into account not only the time variations of the total solar Lyman-$\alpha$ flux with time and the spectral shape of the solar Lyman-$\alpha$ line, but also time- and location-dependent modifications of radiation pressure due to absorption of solar photons by ISN H in the inner heliosphere.

\acknowledgments
The authors wish to thank the anonymous referee of our last paper \citep{kowalska-leszczynska_etal:18a} for drawing our attention to the problem of the absorption of solar radiation by ISN H inside the heliosphere. This study was supported by Polish National Science Center grant 2015-18-M-ST9-00036.

\bibliographystyle{aasjournal}
\bibliography{iplbib}

\end{document}